\documentclass[epj]{svjour}

\usepackage{graphicx}
\usepackage{amsmath}
\usepackage{amssymb}

\newcommand{\be}{\begin{equation}}
\newcommand{\ee}{\end{equation}}
\newcommand{\rmd}{\mathrm{d}}
\newcommand{\kB}{k_\mathrm{B}}

\title{Casimir attractive--repulsive transition in MEMS}

\date{\today}

\author{M. Bostr{\"o}m\inst{1,2}\and S. {\AA}. Ellingsen\inst{1} \and I. Brevik\inst{1}\and M. Dou\inst{2} \and C. Persson\inst{2,3} \and Bo E. Sernelius\inst{4}\thanks{E-mail: bos@ifm.liu.se} }

\institute{
	\inst{1} Department of Energy and Process Engineering, Norwegian University of Science and Technology, \\ N-7491 Trondheim, Norway \\
	\inst{2} Department of Materials Science and Engineering, Royal Institute of Technology, SE-100 44 Stockholm, Sweden, EU\\
	\inst{3} Department of Physics, University of Oslo, P.O.  Box 1048 Blindern, N-0316 Oslo, Norway\\
	\inst{4} Division of Theory and Modeling, Department of Physics,  Chemistry and Biology, Link\"{o}ping University, SE-581 83 Link\"{o}ping, Sweden, EU }

\begin{document}

\abstract{
  Unwanted stiction in micro- and nanomechanical (NEMS/MEMS) systems due to dispersion (van der Waals, or Casimir) forces is a significant hurdle in the fabrication of systems with moving parts on these length scales. Introducing a suitably dielectric liquid in the interspace between bodies has previously been demonstrated to render dispersion forces repulsive, or even to switch sign as a function of separation. Making use of recently available permittivity data calculated by us we show that such a remarkable non-monotonic Casimir force, changing from attractive to repulsive as separation increases, can in fact be observed in systems where constituent materials are in standard NEMS/MEMS use requiring no special or exotic materials. No such nonmonotonic behaviour has been measured to date. We calculate the force between a silica sphere and a flat surface of either zinc oxide or hafnia, two materials which are among the most prominent for practical microelectrical and microoptical devices. Our results explicate the need for highly accurate permittivity functions of the materials involved for frequencies from optical to far-infrared frequencies. A careful analysis of the Casimir interaction is presented, and we show how the change in the sign of the interaction can be understood as a result of multiple crossings of the dielectric functions of the three media involved in a given set-up. 
}

%\pacs{34.20.Cf}{Interatomic potentials and forces }
%\pacs{03.70.+k}{Theory of quantized fields}
%\pacs{68.37.Ps}{Atomic force microscopy (AFM)}

\maketitle

\section{Introduction}
More than 60 years ago Casimir predicted \cite{Casi} that boundary effects on the electromagnetic fluctuations can produce attraction between a pair of parallel, closely spaced, 
perfectly conducting plates.
His calculation was extended to real materials by Lifshitz \cite{lifshitz55}.
Only half a century later was precision measurement of the Casimir-Lifshitz force between macroscopic bodies made possible \cite{Lamo,mohideen98,decca05,Sush}, confirming the theory of Casimir and Lifshitz even though certain discrepancies between theory and experiment still persist \cite{Bost2,brevik06}.
Although in all its most immediate manifestations the Casimir force is attractive, theoretical schemes have long existed whereby Casimir repulsion may be achieved \cite{Dzya,Rich1,Rich,boyer74,milton12}. Munday, Capasso, and Parsegian \cite{Mund} famously demonstrated that the Casimir--Lifshitz force could be repulsive by a suitable choice of interacting surfaces in a fluid, following similar experiments preceding it \cite{rep}. 

We show in the present work that Casimir repulsion, and even transitions from attraction to repulsion with varying separation, are possible with some of the most important materials in use in the fields of micro and nanoelectronics and microoptics. Although the introduction of a dielectric liquid is still required (we suggest bromobenzene as used in \cite{Mund}), no further use of special or exotic materials is necessary. Change of sign of the Casimir force with increasing separation was analyzed by Phan and Viet \cite{Phan11}, and by Bostr{\"o}m {\it et al.} \cite{Bos2012,Bost2012B} and the concept was already familiar from theory and experiments in the context of films on surfaces \cite{Dzya,AndSab,Haux,Ninh}. Here we present calculations making use of two important materials within microelectronics, zinc oxide ($\mathrm{ZnO}$) and hafnia ($\mathrm{HfO}_2$), made possible by newly calculated dielectric data for these materials. In simplistic terms, the necessary requirement for Casimir repulsion is that the permittivities of the two bodies and the interspatial liquid satisfy the inequality
\be
  \varepsilon_\text{body 1} >\varepsilon_\text{interspace} >\varepsilon_\text{body 2}.
\ee
Subtleties such as change of sign occurs because the dielectric response of a medium to an imposed field depends strongly on the field's frequency, so the inequality may be satisfied in some frequency ranges, but not in others. The Casimir force depends on the response over a broad range of frequencies, although the main contributions come from frequencies where $\omega d/c \lesssim 1$ ($d$ is the separation between bodies), which explains how the sign of the force can depend on the body-body separation.

We have calculated the Casimir force between materials which are already important in nano- and microelectromechanical systems (NEMS and MEMS), separated by the oil bromobenzene, to show that these standard materials are sufficient to observe both attraction and repulsion in one and the same set-up. Zinc oxide is a multi-functional semiconductor much used in optoelectronic devices. The 
material is transparent at optical frequencies while blocking the ultraviolet light, and it becomes highly conductive when $n$-type doped. Hafnia is commonly used for optical coating and is a leading candidate for the replacement of silica for a number of microelectronic applications due to its high permittivity, low optical absorption, and low thermal expansion \cite{wilk01}.  

To make contact with experiments \cite{Mund} we consider the geometry consisting of a silica sphere (such as may be attached to an atomic force microscope tip \cite{mohideen98}) interacting with a ZnO or hafnia surface across bromobenzene (Bb). 
Apart from demonstrating the possibility of transition from attractive to repulsive Casimir forces, our calculations
demonstrate the importance of having access to high accuracy dielectric functions \cite{Lam,Svet0,Zwol0}. Different levels of modeling of the dielectric functions give radically different results. For instance, the sign of the force beyond about 50 nanometers may even change depending on whether low frequency electron-phonon contributions are included in the modeling or not. We review the calculations of the dielectric functions of the different media and recapitulate the theory of Casimir-Lifshitz forces, whereupon we present numerical results. 

\section{Calculation of dielectric permittivities}

\begin{figure}
\includegraphics[width=7.8cm]{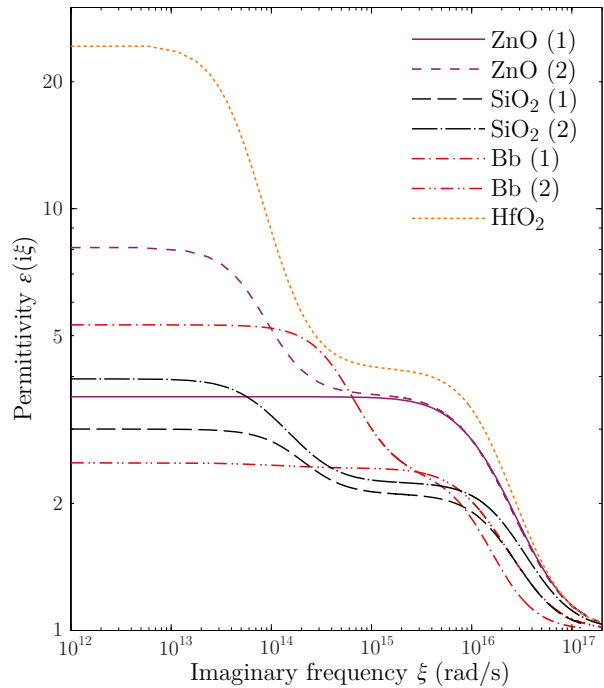}
\caption{(Colour on-line) Dielectric functions at imaginary frequencies for ZnO, ${\rm{Si}}{{\rm{O}}_2}$, Bb (bromobenzene), and HfO$_2$. The results for ZnO and HfO$_2$ are from calculations in 
this work using the $GW_0$ method based on the DFT \cite{Dou,Gonz}; the first ZnO version neglects and the second includes phonon contributions. For ${\rm{Si}}{{\rm{O}}_2}$ the first version is from an oscillator model \cite{Grab}, while the second is the present $GW_0$ results. For Bb the first result \cite{Mund}, is from an oscillator model; the second \cite{Zwol}  from a different oscillator model; note that for the second of the dielectric functions for Bb the static value only entering the $n = 0$ term is 5.37. }
\label{figu1}
\end{figure}

The Casimir force at temperature $T$ can be calculated if the dielectric functions (for discrete imaginary frequencies, $\xi_n = 2\pi n\kB T/\hbar $) are known.
The dielectric functions of the materials play the essential role in the Lifshitz theory \cite{Dzya}. Figure \ref{figu1} presents the  dielectric functions on the imaginary frequency axis for bromobenzene (Bb) determined in \cite{Mund}, SiO$_2$ (both calculated and a modeled dispersion in \cite{Grab} based on experiments), ZnO, and HfO$_2$. For the theoretically determined dielectric functions in this work, we present the calculations both including and excluding the optical phonon modes in order to illustrate the importance of modeling the electron-phonon coupling for analyzing the spectra at frequencies below $\sim {10^{15}}$ rad/s.

The complex dielectric functions for wurtzite ZnO, monoclinic HfO$_2$ and $\alpha$-quartz SiO$_2$ were determined employing a first-principles approach within the density functional theory (DFT).
The electronic structure, neglecting electron-phonon coupling, was obtained from the partial self-consistent $GW_0$ method where the Green functions are updated iteratively whereas the screened Coulomb potential $W$ is fixed \cite{Shishkin,Dou}.
 The imaginary part of the dielectric function was calculated in the long wavelength limit from the joint density of
states and the optical momentum matrix. Since the dielectric function in polar materials can depend strongly on the electron-phonon coupling, we modeled this contribution to the dielectric function using the Lorentz model and Kramers-Heisenberg formula with multi-phonon contribution \cite{Lor,Kuz}. The phonon part of the dielectric function is:
\be
\Delta\varepsilon_\mathrm{phonon}(\omega)=\sum_j \frac{(\omega^2_{\mathrm{LO},j}-\omega^2_{\mathrm{TO},j})\cdot \varepsilon_{\infty,j}}{\omega^2_{\mathrm{TO},j}-\omega^2-i\gamma_j \omega}.
\ee
Here,  $\omega_{\mathrm{LO},j}$ and $\omega_{\mathrm{TO},j}$ are the longitudinal optical (LO) and transverse optical (TO) phonon frequencies of the jth mode, respectively, and ${\varepsilon _{\infty ,j}}$ is the high frequency dielectric constant of the jth phonon mode. We determine $\epsilon_{\infty,j}$ by employing experimental data of
the phonon frequencies \cite{Ash}. The low-energy spectra is verified by
calculating the static dielectric constant from the Born effective
charges.
The calculated dielectric functions of ZnO, SiO$_2$, and HfO$_2$ on the imaginary frequency axis are shown in Fig. \ref{figu1} in the small damping limit (i. e., phonon damping parameters ${\gamma _j} \to 0$ ). The corresponding static dielectric constants are for ZnO 7.9 (8.1-8.3), for SiO$_2$ 3.9 (3.9-4.4), and for HfO$_2$ 24 (15-25), respectively. The numbers within parentheses are the experimental values from Refs. \cite{Lan,Stri,Rob,Ash2}.

\section{The Casimir force formalism}

Using the  Deryaguin (or proximity force) approximation \cite{Der} the Casimir--Lifshitz force of a planar surface of material 1 (ZnO or HfO$_2$) with a sphere (radius $R$) of material 3 (SiO$_2$) across medium 2 (Bb) results in a summation of imaginary frequency terms \cite{Dzya,Ser,Der}:
\begin{equation}
F = 2 \pi R\sum\limits_{n = 0}^\infty {}^{'} {g\left( \xi_n \right)}.
\label{equ1}
\end{equation}
Note that positive values of $F$ correspond to repulsive force.

In the retarded treatment there are contributions from the two light polarisations, transverse magnetic (TM) and transverse electric (TE), $g(\xi_n )= g^\mathrm{TM}(\xi_n) + g^\mathrm{TE}( \xi_n)$, 
where
\begin{align}
g^\mathrm{TM}(\xi _n) =& \kB T\int \frac{\rmd^2\, q}{(2\pi)^2}\notag \\
&\times \ln \left[1 -  r_{21}^\mathrm{TM}(i\xi_n)r_{23}^\mathrm{TM}(i\xi_n)e^{ - 2\gamma _2(i\xi_n)d} \right],\\
g^\mathrm{TE}(\xi _n) =& \kB T\int \frac{\rmd^2\, q}{(2\pi)^2}\notag \\
&\times \ln \left[1 -  r_{21}^\mathrm{TE}(i\xi_n)r_{23}^\mathrm{TE}(i\xi_n)e^{ - 2\gamma _2(i\xi_n)d} \right],\\
\gamma_i(i\xi_n) =& \sqrt{q^2 - \varepsilon_i(i\xi_n)(i\xi_n/c)^2},
\label{equ2}
\end{align}
and the Fresnel reflection coefficients are
\begin{subequations}
\begin{align}
  r_{2i}^\mathrm{TM}(i\xi_n)=&\frac{\varepsilon_2(i\xi_n)\gamma_i(i\xi_n)-\varepsilon_i(i\xi_n)\gamma_2(i\xi_n)}{\varepsilon_2(i\xi_n)\gamma_i(i\xi_n)+\varepsilon_i(i\xi_n)\gamma_2(i\xi_n)},\\
  r_{2i}^\mathrm{TE}(i\xi_n)=&\frac{\gamma_i(i\xi_n)-\gamma_2(i\xi_n)}{\gamma_i(i\xi_n)+\gamma_2(i\xi_n)}.
\end{align}
\end{subequations}
The non-retarded limit can be investigated by letting the velocity of light go to infinity\footnote{In the model case where reflectivity does not tend to zero as $\xi\to \infty$, it is necessary to retain finite speed of light for convergence. For perfectly reflecting surfaces, thus, there is no non-retarded regime.}:
\be
  F_\mathrm{nonret.} \approx  -\frac{R\kB T}{4d^2}\sum\limits_{n = 0}^\infty {}^{'} \mathrm{Li}_3\bigl[r_{21}^\mathrm{TM}(i\xi_n)r_{23}^\mathrm{TM}(i\xi_n)\bigr],
\ee
where the order $3$ polylogarithm is 
\be
  \mathrm{Li}_3(x)=\sum_{n=1}^\infty \frac{x^n}{n^3}.
\ee
The non-retarded approximation is the short separation asymptote of the real (retarded) Casimir force.

Frequency intervals where the intervening medium has a dielectric permittivity in between the permittivities of the two bodies give a repulsive contribution as $r_{21}r_{23}<0$ for both polarisations in these intervals; other intervals give an attractive contribution. Correct calculation of the force --- even its sign --- thus requires accurate dielectric functions for a wide frequency region. Since, roughly speaking, the expression (\ref{equ1}) picks up its main contribution from $\xi\lesssim c/d$, the sign of the force in the short and long separation regimes can be designed by choosing materials with the appropriate dielectric properties in different frequency ranges. In particular, the long-separation range $d \gg c/\xi_1$ is dictated entirely by the $n=0$ term of the sum,
\be\label{farasymp}
{F_{n = 0}} =  - \frac{{R{k_B}T}}{{8{d^2}}}{\rm{L}}{{\rm{i}}_3}\left[ {\frac{{{\varepsilon _2}(0) - {\varepsilon _3}(0)}}{{{\varepsilon _2}(0) + {\varepsilon _3}(0)}}\frac{{{\varepsilon _2}(0) - {\varepsilon _1}(0)}}{{{\varepsilon _2}(0) + {\varepsilon _1}(0)}}} \right].
\ee
Since we are limited by the proximity force approximation to $d \ll R$ this long-separation range is only valid if $R$ is large enough.  In Fig. \ref{figu2} we see that for our choice of materials the $n=0$ asymptote starts to dominate the result at fractions of a micrometer. Thus, for the long-separation range to be of interest here the radius of the sphere has to be of micrometer size or larger. To avoid misunderstandings we repeat that all presented results are valid only for $d \ll R$. For larger separations, $d\gtrsim R$ or larger, all results are different. The $n=0$ term is different and represents the long-separation asymptote that is approached when $d \to \infty $.

\section{Numerical examples}

\begin{figure}
\includegraphics[width=\columnwidth]{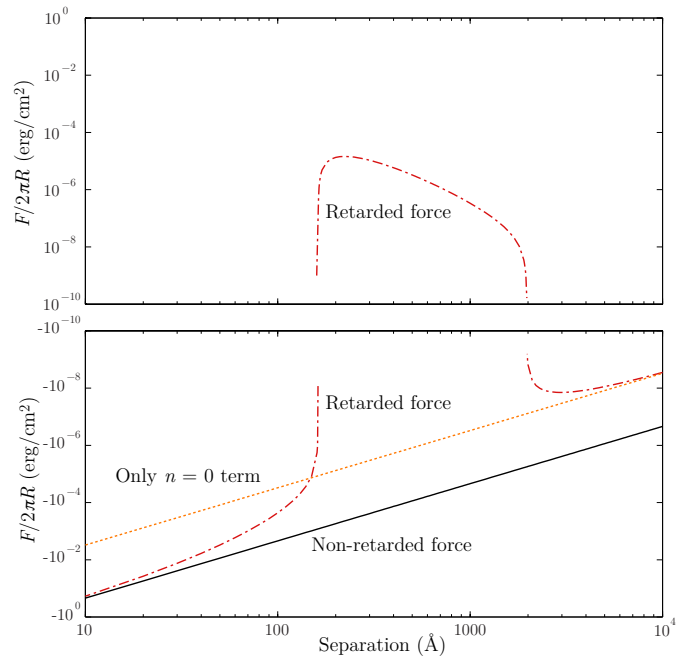}
\caption{(Colour on-line) Casimir-Lifshitz force between ZnO (1) surface and SiO$_2$ (1) sphere in Bb (1).  The fully retarded free energy is attractive in the small and large separation limits, but is repulsive in an intermediate interval. The nonretarded energy and the $n = 0$ contribution are both attractive and fall off with increasing $d$ as $d^{-2}$.}
\label{figu2}
\end{figure}

\begin{figure}
\includegraphics[width=\columnwidth]{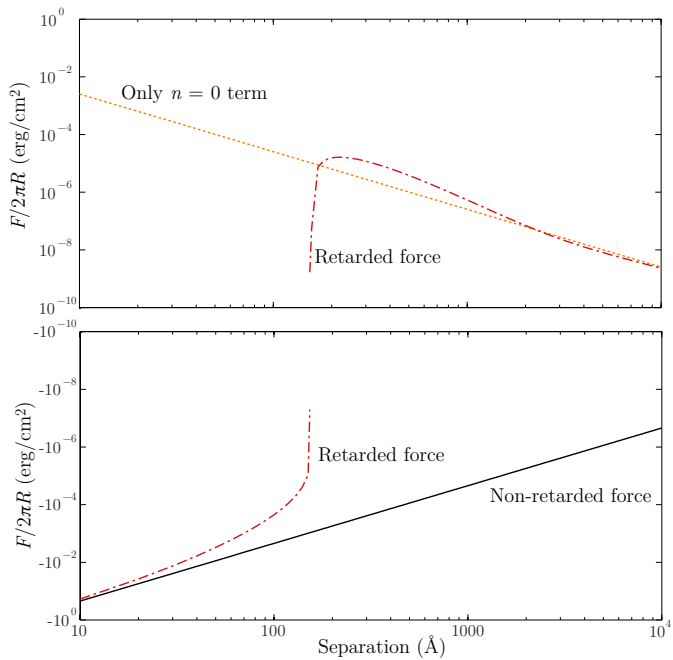}
\caption{(Colour on-line) Casimir-Lifshitz force between ZnO (2) surface and SiO$_2$ (2) sphere in Bb (2). The nonretarded energy is attractive whereas the $n = 0$ contribution is repulsive, whereas the fully retarded (actual) force changes sign at an intermediate separation. Both asymptotes decrease as $ d^{-2}$.}
\label{figu3}
\end{figure}

\begin{figure}
\includegraphics[width=\columnwidth]{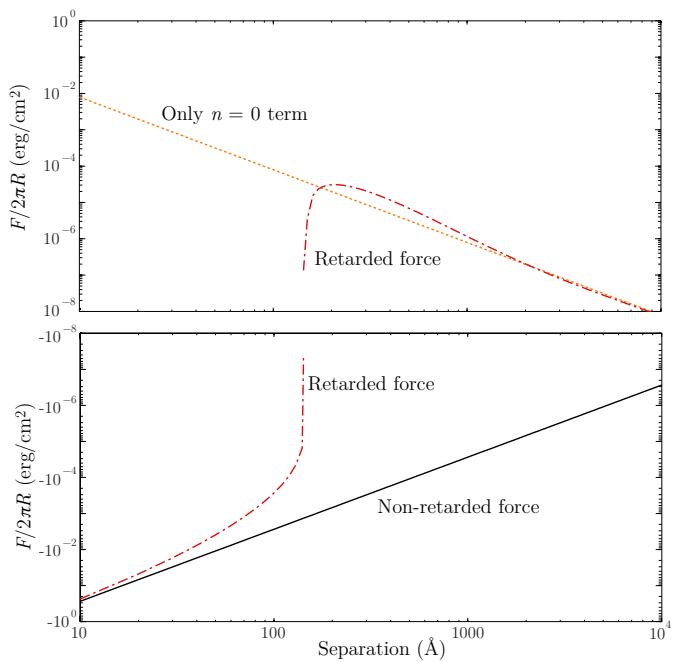}
\caption{(Colour on-line) Casimir-Lifshitz force between HfO$_2$ surface and SiO$_2$ (2) sphere in Bb (2). The nonretarded energy  is attractive while the $n = 0$ contribution is repulsive, and the fully retarded (actual) force tends to these asymptotically in the short and long separation limits, respectively. }
\label{figu4}
\end{figure}

Using the theory and permittivity data laid out in the above sections, numerical calculations are straightforward. All results presented here are for 300 K. We show  in Figs.\,\ref{figu2} and\,\ref{figu3}  the retarded force, nonretarded force, and the n = 0 term. It is clear that retardation changes  in a major way the Casimir-Lifshitz force between a silica sphere and zinc oxide surface in bromobenzene. The figures are in log-log scale so as to visualise the changing sign of the forces.

By means of optical measurements or accurate theoretical calculations of the dielectric properties of the interacting objects and liquid it is possible to predict the force --- from short-range attractive van der Waals force to  intermediate range repulsive Casimir force. The long range entropic asymptote is either attractive or repulsive depending on the optical properties for low frequencies. 

Upon inspection of figures \ref{figu2} and \ref{figu3} it is striking that the two different permittivity models can predict different signs for the large separation asymptote of the Casimir force between silica and ZnO. Mathematically this can be understood from equation (\ref{farasymp}) by regarding the zero frequency limit of the different permittivities involved: the low frequency asymptote of the permittivity of the liquid bromobenzene falls either between or above those of the solids depending on whether or not low frequency electron-phonon contributions are included for ZnO. This highlights that in order to predict the Casimir force at larger separations, indeed even its sign, care must be taken that the model permittivities used take realistic values in the quasistatic limit. Note that the predictions in the two cases differ greatly at all separations above about 100\,nm, where the Casimir effect can still play a practical role in MEMS, thus the drastic effects of having inadequate permittivity data is not just a curiosity but could have important consequences.

Figures \ref{figu3} and \ref{figu4} present our best estimates for the Casimir effect between a silica sphere and, respectively, ZnO and HfO$_2$, in bromobenzene. These calculations draw on the best available dielectric data for ZnO, SiO$_2$ and HfO$_2$ as determined from the $GW_0$ calculations, whereas permittivity data for Bb are taken from \cite{Zwol}. Both of these systems, which apart from the interspatial oil consist only of typical MEMS materials, exhibit similar behaviour: the Casimir force is attractive at short separations and becomes repulsive at larger separations. No such non-monotonic Casimir force has been measured to date, yet our calculations predict that it may be not only observable, but even of practical importance in realistic microelectromechanical set-ups. The proximity force approximation is valid if the sphere radius is much larger than the separation \cite{Ser2,Ser3}. Thus the sphere radius must be in the micro meter range or larger for the rightmost parts of figures \ref{figu2}-\ref{figu4} to be valid.

\section{Conclusions}

We have shown herein that non-monotonic Casimir force, changing from attractive to repulsive with increasing separation, may be found in set-ups in which the constituent materials are in standard use in NEMS and MEMS. To wit, we have calculated the Casimir force between a silica sphere and a half-space of ZnO and HfO$_2$, respectively, immersed in bromobenzene. All three solids are among the primary materials used in NEMS and MEMS. 

The striking change of sign of the Casimir force is a consequence of the frequency dependence of the three different permittivities involved. At small separations, all frequencies contribute to the force, whereas the force at long separations depends on the materials' quasistatic dielectric response only. It is possible therefore for the force to change sign more than once, and the force at long separations, indeed even its sign, depends sensitively on the low frequency asymptotics of the dielectric function model used. It is of vital importance for accurate force calculation that low frequency contributions from the far-infrared regime, such as photon-phonon interactions, are included in the permittivity function. This frequency regime is typically not covered in tables of optical data, but its exclusion could cause radically wrong predictions of the Casimir force, not only in the long-separation asymptote, but also at transitional separations between the short and long distance limits. 

We suggest, in conclusion, that measurements of the attraction--repulsion transition of the Casimir force may not only be possible, but of considerable practical interest. In our example systems, seen in figures \ref{figu3} and \ref{figu4}, the transition happens at about 150\,nm. This transition separation can be modified by slighly altering the dielectric properties involved. It is straightforward to generalize the results, for example, by using ultra-thin coatings of, e.g., molybdenum disulfide or graphene \cite{bosser2012,Svet}. Such surface modifications may alter the way Casimir-Lifshitz forces switch from attraction to repulsion to attraction/repulsion.

\begin{acknowledgement}
M.B.  acknowledges support from an ESF exchange grant within the activity ``New Trends and Applications of the Casimir Effect", through the network CASIMIR. C.P. and M. B. acknowledge support from VR (Contract No. 90499401) and STEM (Contract No. 34138-1). B.E.S. acknowledges financial support from VR (Contract No. 70529001). 
\end{acknowledgement}

%SAE: changed all refs to EPL style
%BES: changed all refs to EPJ style

\end{document}